\documentclass[final,5p,times,twocolumn]{elsarticle}
\usepackage{graphicx}
\usepackage{amsmath,amssymb}
\newcommand{\superlat}{\sqrt{3}\times\sqrt{3}R30^\circ}
\newcommand{\ekd}[1]{e^{i\vec{k}\cdot\vec{\delta}_#1}}
\journal{Physica E}
\begin{document}
\begin{frontmatter}
\title{Uniaxial strain on gapped graphene}
\author[ipm]{M. Farjam\corref{cor1}}
\ead{mfarjam@mail.ipm.ir}
\cortext[cor1]{Corresponding author}
\author[ipm,sbu]{H. Rafii-Tabar}
\address[ipm]{Department of Nano-Science, Institute for Research in
Fundamental Sciences (IPM), P.O. Box 19395-5531, Tehran, Iran}
\address[sbu]{Department of Medical Physics and Biomedical Engineering, and
Research Centre for Medical Nanotechnology and Tissue Engineering, Shahid
Beheshti University of Medical Sciences, Evin, Tehran 19839, Iran}

\begin{abstract}
We study the effect of uniaxial strain on the electronic band structure of
gapped graphene. We consider two types of gapped graphene, one which breaks the
symmetry between the two triangular sublattices (staggered model), and another
which alternates the bonds on the honeycomb lattice (Kekul\'e model). In the
staggered model, the effect of strains below a critical value is only a shift of
the band gap location. In the Kekul\'e model, as strain is increased, band gap
location is initially pinned to a corner of the Brillouin zone while its width
diminishes, and after gap closure the location of the contact point begins to
shift. Analytic and numerical results are obtained for both the tight-binding
and Dirac fermion descriptions of gapped graphene.
\end{abstract}


\end{frontmatter}

\section{Introduction}
\label{}

Recently, strain engineering of the electronic structure has been explored as an
alternative method in the design of graphene-based electronic circuitry
\cite{pereira2009a}. The approach is based on generating local strains to change
the hopping amplitudes in an anisotropic way, which in turn leads to the
presence of effective gauge fields for the Dirac electrons  \cite{neto2009}. A
key finding is that for small and moderate uniaxial deformations the gapless
Dirac spectrum is robust, and a gap opens only for large deformations above a
particular threshold \cite{pereira2009b,ni2009,farjam2009b}. More precisely,
the presence of anisotropy in the tight-binding hoppings on a honeycomb lattice
makes the Dirac points approach each other until they merge at a critical
asymmetry, at which point a band gap begins to open
\cite{hasegawa2006,goerbig2008,montambaux2009,wunsch2008}.

For practical purposes, graphene can be considered as a gapless semiconductor.
Nevertheless, gaps of various origins have been identified in graphene.
Intrinsically, spin-orbit coupling is responsible for a tiny gap, on the order
of $10^{-3}$~meV \cite{kane2005,min2006,huertas2006,yao2007}, and
electron-electron interactions may render graphene an insulator in vacuum
\cite{drut2009}. Extrinsically, interaction with substrates and adlayers can
induce a band gap in graphene.  Epitaxial graphene has a band gap
\cite{zhou2007} which has been explained as substrate-induced \cite{kim2008}.
Other substrates studied include boron-nitride and Cu \cite{giovannetti2007},
Ni \cite{gruneis2008}, and several other metals \cite{khomyakov2009}.  Band gaps
induced by the adsorption of water and ammonia molecules \cite{ribeiro2008}, and
alkali metals \cite{farjam2009a} have been studied theoretically.  Generally,
there are two ways of inducing a gap in monolayer graphene.  One way is the
mixing of electronic states with different pseudospins in the same valley, and
another way is the mixing of states that belong to different valleys. The former
can be achieved by sublattice symmetry breaking which leaves the $A$ and $B$
carbon atoms in different environments, while the latter is produced by certain
translational symmetry breakings \cite{farjam2009a}.

The purpose of the present work is to study the effect of uniaxial strain on the
band structure of gapped graphene \cite{semenoff2008}. As we show, the effect
depends on the origin of the band gap. If it is induced by $AB$ sublattice
symmetry breaking, small and moderate strains do not change the width of the
gap, but cause its locations in $k$-space to move in a similar way as those of
the corresponding Dirac points in the gapless case. On the other hand, if the
band gap is induced by translational symmetry breaking that couples the
different valley states, a different behavior emerges.  With increasing uniaxial
strain, the band gap first diminishes and closes in its fixed location and,
afterwards, the neutrality point is shifted in reciprocal lattice as in the
gapless case.

Our paper is organized as follows.  Section \ref{sec:tbm} contains our analysis
of the relevant tight-binding models. Section \ref{sec:de} contains analysis of
the corresponding Dirac equations. Section \ref{sec:nr} contains numerical
results and their discussion, and Section \ref{sec:con} presents our
conclusions.

\section{Tight-binding model} \label{sec:tbm}

\begin{figure}
\includegraphics[width=8cm]{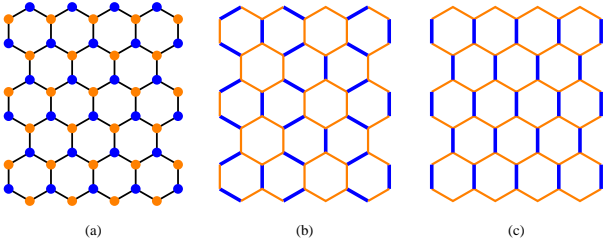}
\caption{\label{fig1} (a) Staggered, (b) Kekul\'e, and (c) quinoid models.}
\end{figure}

The models we use can be described as different types of strain distributions
\cite{saito1998} shown in Fig.~\ref{fig1}. Sublattice symmetry breaking, shown
in Fig.~\ref{fig1}(a), can be associated with an out-of-plane strain
distribution.  The so-called Kekul\'e distortion, shown in Fig.~\ref{fig1}(b),
breaks the translational symmetry in a way which can be described by a
$\superlat$ commensurate lattice, and can be the result of an in-plane strain
distribution.  Finally, the quinoid distortion of graphene, shown in
Fig.~\ref{fig1}(c), represents in-plane uniaxial strain parallel to the
nearest-neighbor bond in the vertical direction.  Our tight-binding models based
on this description are defined by on-site energies and the nearest-neighbor
hoppings, but we neglect the change in bond lengths and use the perfect
honeycomb lattice.

\begin{figure}
\includegraphics[width=8cm]{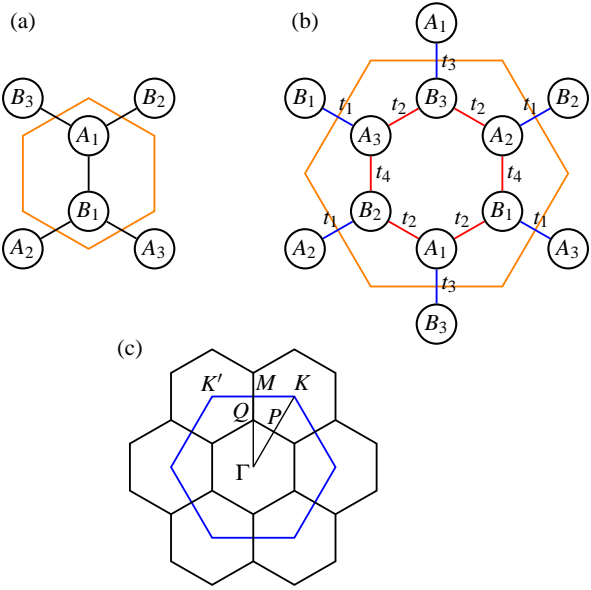}
\caption{\label{fig2} (a) Primitive unit cell of graphene. The
vectors $\vec{\delta}$ defined as vectors from atom $A_1$ to its nearest
neighbors on the $B$ sublattice are given by $\vec{\delta}_1=(0,-1/\sqrt{3})a$,
$\vec{\delta}_2=(1/2,1/2\sqrt{3})a$ and $\vec{\delta}_3=(-1/2,1/2\sqrt{3})a$.
(b) Kekul\'e-type graphene primitive unit cell containing its six atom basis.
The Kekul\'e distortion is characterized by two kinds of hoppings, one set
around the hexagonal ring (red), and the other set crossing the unit cell
boundaries (blue). The addition of a uniaxial strain in the vertical direction
results in four hoppings, $t_1$, $t_2$, $t_3$ and $t_4$. (c) The relationship
between the two Brillouin zones. The path $\Gamma{K}M\Gamma$ is used in the
band plots.}
\end{figure}

The primitive unit cell of graphene and the nearest neighbors of the $A$ and $B$
atoms in the unit cell are shown in Fig.~\ref{fig2}(a).  The tight-binding
models involve the vectors $\vec{\delta}$ from an $A$ atom to its
nearest-nearest neighbors which are defined in the caption of Fig.~\ref{fig2}.
The primitive unit cell of the Kekul\'e model is shown in Fig.~\ref{fig2}(b).
The Brillouin zone of the Kekul\'e model and that of graphene, the larger
hexagon, are depicted in Fig.~\ref{fig2}(c), with special symmetry points
defined. In the Kekul\'e model the points $K$, $K'$ and $\Gamma$ belong to the
reciprocal lattice and are therefore equivalent, which results in the coupling
of the two inequivalent valleys that is responsible for the band gap. 

A few parameters are needed to define the tight-binding models.  In addition to
the hopping parameter $t$ ($\approx-2.7$~eV for graphene), there is the band gap
of gapped graphene and the asymmetry in hoppings caused by uniaxial strain.  We
denote by $\Delta$ the half-width of the band gap, and by $\delta{t}$ the shift
in one of the hoppings which is in the direction of strain axis. The parameter
$t$ may be taken to be positive without loss of generality, and it can serve as
the energy scale, so that the models contain two adjustable parameters
$\Delta/t$ and $\delta{t}/t$.

The gapped graphene represented by Fig.~\ref{fig1}(a) can be described by the
tight-binding model with staggered on-site energies $\pm\Delta$.  Including the
effect of uniaxial strain, we can write the Hamiltonian of this
\textit{staggered model} as

\begin{equation} \label{eq:sm}
H_s(\vec{k})=\left( \begin{matrix}
-\Delta & h(\vec{k}) \\
h^\ast(\vec{k}) & \Delta
\end{matrix} \right),
\end{equation}
where
\begin{equation}
h(\vec{k})=(t+\delta{t})\ekd{1}+t\ekd{2}+t\ekd{3}.
\end{equation}
Diagonalizing the Hamiltonian defined by Eq.~(\ref{eq:sm}), we find the energy
bands,
\begin{equation} \label{eq:smev}
\epsilon_{\pm}(\vec{k})=\pm\sqrt{\Delta^2+|h(\vec{k})|^2},
\end{equation}
which show that the band gap is $\ge2\Delta$.  The function $h(\vec{k})$ is the
same as in gapless graphene under strain, where zero modes exist for
$\delta{t}/t<1$ and are shifted from $K$ and $K'$ toward $M$ \cite{note}.
Setting $h(\vec{k})=0$, and taking $k_ya=2\pi/\sqrt{3}$ and $k_xa=2\pi/3+p_xa$,
to move along the $KK'$ line, we find the shift from $K$ to be
\begin{equation} \label{eq:shift}
p_x=\frac{2}{a}\cos^{-1}\left[\frac{1}{2}\left(1+\frac{\delta{t}}{t}\right)
\right]-\frac{2\pi}{3a}.
\end{equation}

We now turn to the gapped graphene based on Kekul\'e distortion, shown in
Fig.~\ref{fig1}(b), which has zero on-site energies but two alternating
hoppings, on one-third and two-thirds of the bonds, respectively
\cite{farjam2009a,hou2007},
\begin{equation} \label{eq:kekule}
t_1=t+\frac{2}{3}\Delta, \qquad
t_2=t-\frac{1}{3}\Delta,
\end{equation}
where, as in the staggered model, $\Delta$ is half of the energy gap.  When
uniaxial strain is applied to the \textit{Kekul\'e model}, we assume that a
shift of $\delta{t}$ is added to the hoppings that are parallel to the strain
axis. We define these hoppings as
\begin{equation}
t_3=t_1+\delta{t}, \qquad t_4=t_2+\delta{t}.
\end{equation}

The Hamiltonian of the Kekul\'e model is given by a $6\times6$ matrix,
\begin{equation} \label{eq:hk}
H_K(\vec{k})=\left( \begin{matrix}
\mathbf{0}_{3\times3} & H_{AB}(\vec{k}) \\
H_{BA}(\vec{k}) & \mathbf{0}_{3\times3} \end{matrix} \right),
\end{equation}
where the $3\times3$ block $H_{AB}$ is
\begin{equation} \label{eq:block}
H_{AB}(\vec{k})=\left( \begin{matrix}
t_2\ekd{2} & t_2\ekd{3} & t_3\ekd{1} \\
t_4\ekd{1} & t_1\ekd{2} & t_2\ekd{3} \\
t_1\ekd{3} & t_4\ekd{1} & t_2\ekd{2} \end{matrix} \right),
\end{equation}
and $H_{BA}(\vec{k})=H^\dag_{AB}(\vec{k})$. The matrix elements of
Eq.~(\ref{eq:block}) have the general form
$t_{ij}\exp(i\vec{k}\cdot\vec{\delta}_{ij})$ when linking $A_i$ and $B_j$ atoms,
and can be read off Fig.~\ref{fig2}(b).

The band structure of the Kekul\'e model can be obtained by numerical
calculation of the eigenvalues of Eq.~(\ref{eq:hk}). However, the existence of
zero modes and their locations are determined more simply by
\begin{equation} \label{eq:det}
\det[H_K(\vec{k})]=\left|\det[H_{AB}(\vec{k})]\right|^2=0.
\end{equation}
Since the zero modes are expected to occur on the horizontal line through
$\Gamma$, we calculate the determinant of $H_{AB}$ for $k_y=0$,
\begin{equation} \label{eq:D}
D(k_x;\Delta/t,\delta{t}/t)=
2\,t_1\,t_2^2 \cos\frac{3k_xa}{2} + t_3(t_4^2 - t_1^2) - 2\,t_4\,t_2^2.
\end{equation}
Setting $D=0$, we find
\begin{equation} \label{eq:kx}
\cos\frac{3k_xa}{2}=\frac{2\,t_4\,t_2^2-t_3(t_4^2 - t_1^2)}{2\,t_1\,t_2^2},
\end{equation}
which has a solution for $k_x$ if the right-hand side is in the interval
$[-1,1]$. In particular, if $\delta{t}=\Delta$ then $t_1=t_4$ and the right-hand
side of Eq.~(\ref{eq:kx}) becomes unity yielding $k_x=0$ which is the $\Gamma$
point, or its equivalent $K$ points.  Therefore, the gap in the spectrum at $K$
vanishes as $\delta{t}$ is increased from $0$ to $\Delta$. Further increase of
$\delta{t}$ then causes the contact points to shift away from the $K$ points,
until they merge at $M$, when the right-hand side of Eq.~(\ref{eq:kx}) becomes
$-1$ and $k_xa=2\pi/3$.

\section{Dirac equation} \label{sec:de}
We can use the Dirac equation to describe the effect of strain on low-energy
electrons provided that both $\Delta/t$ and
$\delta{t}/t\ll1$. The Dirac equation can be derived from the tight-binding
model by setting $\vec{k}=\vec{K}+\vec{p}$ near the $K$ and $K'$ points.  For
the staggered model under strain we obtain
\begin{equation} \label{eq:sd}
\mathcal{H}_s=\left( \begin{matrix}
-\Delta & p^\ast+\delta t & 0 & 0 \\
p+\delta t & \Delta & 0 & 0 \\
0 & 0 & -\Delta & -p+\delta t \\
0 & 0 & -p^\ast+\delta t & \Delta \end{matrix} \right),
\end{equation} 
where $p=p_x+ip_y$ and we have used units such that $\hbar=1$ and
$v_F=ta\sqrt{3}/2=1$. However, we restore $v_F$ explicitly in some of the
derived results below. Here we have followed the convention \[
\psi=[\Phi_{KA},\Phi_{KB},\Phi_{K'A},\Phi_{K'B}]^T \] for the four-dimensional
spinor \cite{manes2007}. Our results can be extended to the band gap due to
spin-orbit interaction which can be described with a similar Hamiltonian as
Eq.~(\ref{eq:sd}), except that the gaps have opposite signs for $K$ and $K'$
points \cite{kane2005}. The energy dispersions derived from Eq.~(\ref{eq:sd})
are given by
\begin{equation} \label{eq:sdev}
\epsilon_{\pm}(\vec{p})=\pm\sqrt{(\pm p_x+\delta t)^2+p_y^2+\Delta^2},
\end{equation}
and can be easily verified to be the limiting cases of Eq.~(\ref{eq:smev}). The
shifts of the Dirac points are $p_x=\mp\delta{t}/v_F$, which agree with
Eq.~(\ref{eq:shift}) for $\delta{t}/t\ll1$.

For the Kekul\'e model under strain we have
\begin{equation} \label{eq:kd}
\mathcal{H}_K=\left( \begin{matrix}
0 & p^\ast+\delta t & 0 & \Delta \\
p+\delta t & 0 & \Delta & 0 \\
0 & \Delta & 0 & -p+\delta t \\
\Delta & 0 & -p^\ast+\delta t & 0 \end{matrix} \right),
\end{equation} 
and the energy eigenvalues are given by
\begin{equation} \label{eq:Kev}
\epsilon_{\pm}(\vec{p})=\pm\left[
\left(\sqrt{p_x^2+\Delta^2}\pm\delta{t}\right)^2+p_y^2\right]^{1/2}.
\end{equation}
For $\delta{t}<\Delta$ there is a gap of size $\Delta-\delta{t}$ at $p=0$. If
$\delta{t}\ge\Delta$, zero modes exist at
\begin{equation}
p_y=0, \qquad p_x=\pm\sqrt{\delta{t}^2-\Delta^2}.
\end{equation}

For $\delta{t}=0$ the energy dispersions, (\ref{eq:sdev}) and (\ref{eq:Kev}),
give identically gapped Dirac spectra, with a density of states (DOS) per unit
area given by
\begin{equation}
\rho(\epsilon)=\begin{cases}
0, &|\epsilon|<\Delta \\
2|\epsilon|/\pi{v_F^2}, &\text{otherwise}.
\end{cases}
\end{equation}
However, for $\delta{t}\ne0$ the DOS of the staggered model remains the same,
while that of the Kekul\'e model changes as the gap shrinks. For
$\delta{t}=\Delta$ and $\epsilon\ll\Delta$, Eq.~(\ref{eq:Kev}) becomes
\begin{equation} \label{eq:esq}
\epsilon^2=\frac{p_x^4}{4\Delta^2}+p_y^2,
\end{equation}
which yields a DOS given by
\begin{equation}
\rho(\epsilon)=\frac{2\Gamma(1/4)}{\pi^{3/2}\Gamma(3/4)}
\frac{\sqrt{\Delta|\epsilon|}}{v_F^2}.
\end{equation}

\section{Numerical results and discussion} \label{sec:nr}
\begin{figure}
\includegraphics[width=8cm]{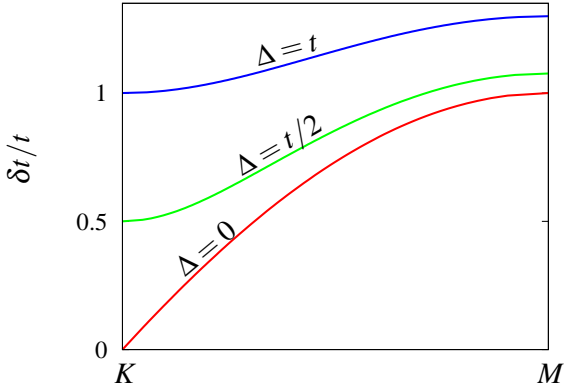}
\caption{\label{fig3} Location of the Dirac point in the
Kekul\'e model along the $KM$ line as a function of $\delta{t}$ for
$\Delta=0,t/2,t$. At $M$ the values of $\delta{t}/t$ are $1,\,1.0763,\,1.2996$,
respectively.}
\end{figure}

Equation (\ref{eq:kx}) can be solved numerically for $k_x$ as a function of
$\Delta$ and $\delta{t}$. The results for $\Delta=0,t/2,t$ are shown in
Fig.~\ref{fig3}. For $\Delta=0$ the same curve can be obtained from
Eq.~(\ref{eq:shift}). For nonzero $\Delta$, the gap first diminishes at $K$ as
$\delta{t}$ is increased from $0$ to $\Delta$, and with further increase of
$\delta{t}$ the Dirac point moves toward $M$.

\begin{figure}
\includegraphics[width=8cm]{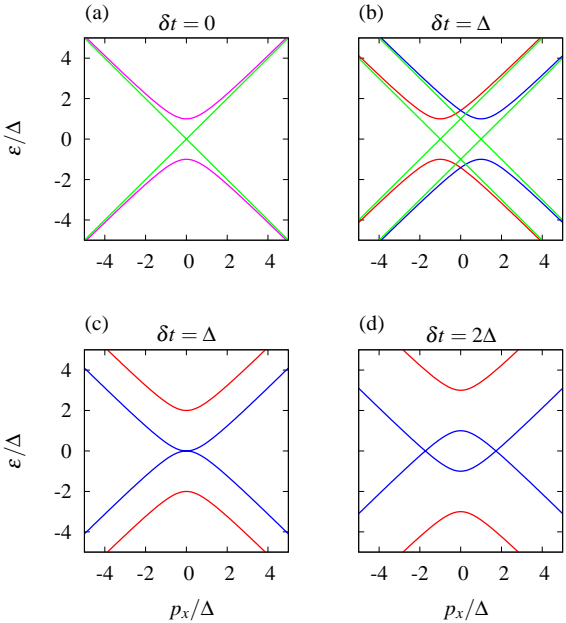}
\caption{\label{fig4} Dirac fermion dispersions along the $p_y=0$ line. (a)
Gapless and gapped Dirac fermions, with gap equal to $2\Delta$. (b) Strain of
$\delta{t}=\Delta$ on gapless and staggered models. (c) Strain of
$\delta{t}=\Delta$ on Kekul\'e model. (d) Strain of $\delta{t}=2\Delta$ on
Kekul\'e model.}
\end{figure}

In Fig.~\ref{fig4} we make plots of Eqs.~(\ref{eq:sdev}) and (\ref{eq:Kev}) for
a few values of $\delta{t}$, using $\Delta$ as the scale of energy and momentum.
For $\delta{t}=0$, shown in Fig.~\ref{fig4}(a), both the staggered and Kekul\'e
models give the same gapped spectrum. However, for $\delta{t}=\Delta$, the
valence and conduction bands shift laterally in the $p_x$ direction without a
change in the gap for the staggered model as in Fig.~\ref{fig4}(b). In contrast,
for $\delta{t}=\Delta$ the gap closes in the Kekul\'e model for one pair of
bands while the other pair are repelled by $2\Delta$, as in Fig.~\ref{fig4}(c).
We note that the dispersion is quadratic instead of linear near the degeneracy
point as can be expected from Eq.~(\ref{eq:esq}).  Further increase of
$\delta{t}$ to $2\Delta$, shown in Fig.~\ref{fig4}(d), causes a shift of the
neutrality points in the $p_x$ direction, and the dispersions become linear
where the bands cross.

\begin{figure}
\includegraphics[width=8cm]{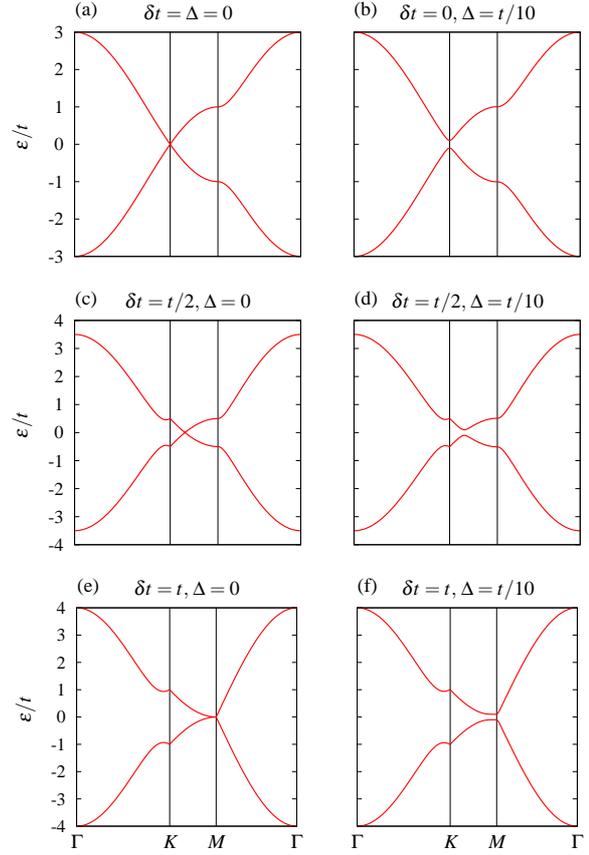}
\caption{\label{fig5} Band structures from the tight-binding model. (a,c,e) show
the band structures of graphene for $\delta{t}=0,\,t/2,\,t$, and
(b,d,f) show the band structures for the staggered model for the same set of
strains.}
\end{figure}

Figure~\ref{fig5} shows the band structures from Eq.~(\ref{eq:smev}) for
$\Delta=0$ and $t/10$. For $\delta{t}=0$, Figs.~\ref{fig5}(a) and (b), the Dirac
point and the gap, respectively, are located at the $K$ point. For
$\delta{t}=t/2$, Figs.~\ref{fig5}(c) and (d), they are shifted by the same
amount to somewhere along the $KM$ line. For $\delta{t}=t$, Figs.~\ref{fig5}(e)
and (f), the shift reaches the $M$ point. The last cases are critical in that
the Dirac points merge at $M$. We can see that the dispersions are quadratic
near $M$ on the $KM$ line, but linear on the $M\Gamma$ line, i.e., a flattening
of the Dirac cones takes place which can be seen in contour plots of the band
structure (as shown in Ref.~6).  For $\delta{t}>1$ a gap opens at the $M$ point
for graphene, and the gap of the staggered model becomes wider.

\begin{figure}
\includegraphics[width=8cm]{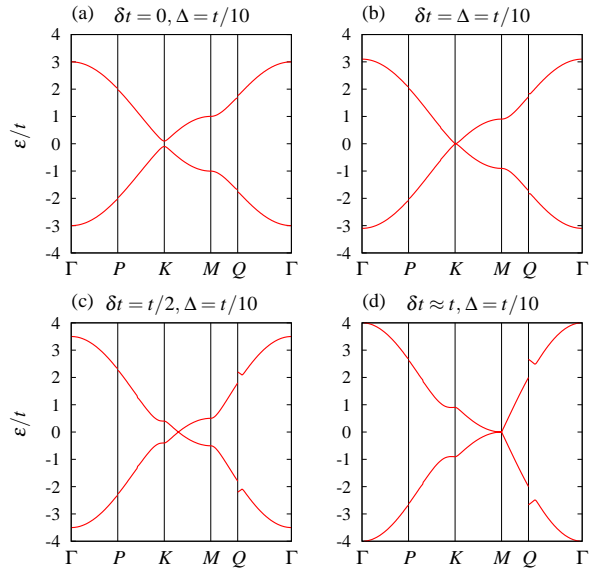}
\caption{\label{fig6} Band structures of the Kekul\'e tight-binding model for
$\delta{t}=0,\,t/10,\,t/2,\,1.00299t$. The extended zone scheme is used.}
\end{figure}

Figure~\ref{fig6} shows the band structures for the Kekul\'e model obtained from
the numerical evaluation of the eigenvalues of Eq.~(\ref{eq:hk}). The band
structures are plotted in the extended scheme so they can be compared with those
of Fig.~\ref{fig5}. The $\Gamma{P}KMQ\Gamma$ path is implied in
Fig.~\ref{fig2}(c), where $P$ and $Q$ are the points where the path crosses the
Brillouin zone of the Kekul\'e model.  It must be remarked that this path does
not enclose the irreducible wedge of the Brillouin zone for strained graphene
\cite{farjam2009b,gui2008}, but includes the $KM$ line where the band crossing
may occur.  Figure~\ref{fig6}(a) may be compared with Fig.~\ref{fig5}(b) which
shows the band structure of the staggered model for the same $\Delta$. The
regions of the gap at the $K$ point are similar in both figures, consistent with
the Dirac equation description, but there are extra gaps in the band structure
of the Kekul\'e model at $P$ and $Q$. As we have seen in Fig.~\ref{fig4}, the
gap closes when $\delta{t}=\Delta$ and this can also be seen in
Fig.~\ref{fig6}(b).  Here the dispersion at $K$ is linear on the $PK$ line and
quadratic on the $KM$ line. A comparison of Figs.~\ref{fig6}(c) and (d) with
Fig.~\ref{fig5}(c) and (e) shows that in the Kekul\'e model, after the gap
closes, the behavior of the neutrality points are not very different from that
in the gapless case.

\section{Conclusions} \label{sec:con}
We considered two models of gapped graphene, denoted by staggered and Kekul\'e,
respectively, and studied the effect of strain on their band structures. We
found that in the staggered model the width of the band gap does not change for
strains less than a critical value, but its locations move following the motion
of the Dirac cones in the gapless case. The effect of strain on the band
structure of the Kekul\'e model is less trivial. With increasing strain,
reflected in changes in the hoppings, the gap begins to diminish with its
locations pinned to the $K$ points. The gap closes when $\delta{t}=\Delta$,
i.e., when the shift in hoppings due to strain equals the half-width of the
original gap and, afterwards, increasing the strain makes the neutrality point
to move in $k$ space as in the gapless case. The effects we have discussed may
be observed experimentally in gapped graphene on Ni substrate and epitaxial
graphene on SiC, respectively.

\section{Acknowledgments}
M.F. acknowledges funding from the Iranian Nanotechnology Initiative and H.R.-T.
from the Iran National Science Foundation.

%\bibliographystyle{elsarticle-num}
%\bibliography{dirac}

\section*{References}

\begin{thebibliography}{10}
\expandafter\ifx\csname url\endcsname\relax
  \def\url#1{\texttt{#1}}\fi
\expandafter\ifx\csname urlprefix\endcsname\relax\def\urlprefix{URL }\fi
\expandafter\ifx\csname href\endcsname\relax
  \def\href#1#2{#2} \def\path#1{#1}\fi

\bibitem{pereira2009a}
V.~M. Pereira, A.~H. {Castro Neto}, Phys. Rev. Lett. 103 (2009) 046801.

\bibitem{neto2009}
A.~H. {Castro Neto}, F.~Guinea, N.~M.~R. Peres, K.~S. Novoselov, A.~K. Geim,
  Rev. Mod. Phys. 81 (2009) 109.

\bibitem{pereira2009b}
V.~M. Pereira, A.~H. {Castro Neto}, N.~M.~R. Peres, Phys. Rev. B 80 (2009)
  045401.

\bibitem{ni2009}
Z.~H. Ni, T.~Yu, Y.~H. Lu, Y.~Y. Wang, Y.~P. Feng, Z.~X. Shen, ACS Nano 3
  (2009) 483.

\bibitem{farjam2009b}
M.~Farjam, H.~Rafii-Tabar, Phys. Rev. B 80 (2009) 167401.

\bibitem{hasegawa2006}
Y.~Hasegawa, R.~Konno, H.~Nakano, M.~Kohmoto, Phys. Rev. B 74 (2006) 033413.

\bibitem{goerbig2008}
M.~O. Goerbig, J.-N. Fuchs, G.~Montambaux, F.~Pi\'echon, Phys. Rev. B 78 (2008)
  045415.

\bibitem{montambaux2009}
G.~Montambaux, F.~Pi\'echon, J.-N. Fuchs, M.~O. Goerbig, Phys. Rev. B 80 (2009)
  153412.

\bibitem{wunsch2008}
B.~Wunsch, F.~Guinea, F.~Sols, New J. Phys. 10 (2008) 103027.

\bibitem{kane2005}
C.~L. Kane, E.~J. Mele, Phys. Rev. Lett. 95 (2005) 226801.

\bibitem{min2006}
H.~Min, J.~E. Hill, N.~A. Sinitsyn, B.~R. Sahu, L.~Kleinman, A.~H. MacDonald,
  Phys. Rev. B 74 (2006) 165310.

\bibitem{huertas2006}
D.~Huertas-Hernando, F.~Guinea, A.~Brataas, Phys. Rev. B 74 (2006) 155426.

\bibitem{yao2007}
Y.~Yao, F.~Ye, X.-L. Qi, S.-C. Zhang, Z.~Fang, Phys. Rev. B 75 (2007)
  041401(R).

\bibitem{drut2009}
J.~E. Drut, T.~A. L\"ahde, Phys. Rev. Lett. 102 (2009) 026802.

\bibitem{zhou2007}
S.~Y. Zhou, G.-H. Gweon, A.~V. Fedorov, P.~N. First, W.~A. de~Heer, D.-H. Lee,
  F.~Guinea, A.~H. {Castro Neto}, A.~Lanzara, Nat. Mater. 6 (2007) 770.

\bibitem{kim2008}
S.~Kim, J.~Ihm, H.~J. Choi, Y.-W. Son, Phys. Rev. Lett. 100 (2008) 176802.

\bibitem{giovannetti2007}
G.~Giovannetti, P.~A. Khomyakov, G.~Brocks, P.~J. Kelly, J.~{van den Brink},
  Phys. Rev. B 76 (2007) 073103.

\bibitem{gruneis2008}
A.~Gr\"uneis, D.~V. Vyalikh, Phys. Rev. B 77 (2008) 193401.

\bibitem{khomyakov2009}
P.~A. Khomyakov, G.~Giovannetti, P.~C. Rusu, G.~Brocks, J.~{van den Brink},
  P.~J. Kelly, Phys. Rev. B 79 (2009) 195425.

\bibitem{ribeiro2008}
R.~M. Ribeiro, N.~M.~R. Peres, J.~Coutinho, P.~R. Briddon, Phys. Rev. B 78
  (2008) 075442.

\bibitem{farjam2009a}
M.~Farjam, H.~Rafii-Tabar, Phys. Rev. B 79 (2009) 045417.

\bibitem{semenoff2008}
G.~W. Semenoff, V.~Semenoff, F.~Zhou, Phys. Rev. Lett. 101 (2008) 087204.

\bibitem{saito1998}
R.~Saito, G.~Dresselhaus, M.~S. Dresselhaus, Physical Properties of Carbon
  Nanotubes, Imperial College Press, 1998.

\bibitem{note}
We have limited our discussion to the case of enhancement of the magnitude of
hopping. One difference is that when the hopping is reduced the Dirac points
move in the opposite direction away from the $M$ point, which can be shown by
making additional band plots along a different path. However, the main reason
to choose enhancement over reduction, is that in the latter the hopping vanishes
before the Dirac points can merge and, therefore, this critical region cannot be
reached. In real experiments, of course, there are even more severe limitations
on how much strain can change the hoppings.

\bibitem{hou2007}
C.-Y. Hou, C.~Chamon, C.~Mudry, Phys. Rev. Lett. 98 (2007) 186809.

\bibitem{manes2007}
J.~L. {Ma\~nes}, F.~Guinea, M.~A.~H. Vozmediano, Phys. Rev. B 75 (2007) 155424.

\bibitem{gui2008}
G.~Gui, J.~Li, J.~Zhong, Phys. Rev. B 78 (2008) 075435.

\end{thebibliography}

\end{document}